# CHROMOSPHERIC INVERSIONS OF A MICRO-FLARING REGION

A. Reid[1,2], V. Henriques[1,3], M. Mathioudakis[1], J. G. Doyle[2], T. Ray[4]

1. Astrophysics Research Centre, School of Mathematics and Physics, Queen's University Belfast, BT7 1NN, Northern Ireland, UK; e-mail:
aaron.reid@qub.ac.uk
2. Armagh Observatory and Planetarium, College Hill, Armagh, BT61 9DG, UK
3. Institute of Theoretical Astrophysics, University of Oslo, PO Box 1029 Blindern, 0315 Oslo, Norway
4. Dublin Institute for Advanced Studies, 31 Fitzwilliam Place, Dublin 2, Ireland

## ABSTRACT

We use spectropolarimetric observations of the Ca II 8542 Å line, taken from the Swedish 1-m Solar Telescope (SST), in an attempt to recover dynamic activity in a micro-flaring region near a sunspot via inversions. These inversions show localized mean temperature enhancements of ~1000 K in the chromosphere and upper photosphere, along with co-spatial bi-directional Doppler shifting of 5 - 10 km s$^{-1}$. This heating also extends along a nearby chromospheric fibril, co-spatial to 10 - 15 km s$^{-1}$ down-flows. Strong magnetic flux cancellation is also apparent in one of the footpoints, concentrated in the chromosphere. This event more closely resembles that of an Ellerman Bomb (EB), though placed slightly higher in the atmosphere than is typically observed.

## 1. INTRODUCTION

Micro-flares are considered as events with energies of ~ $10^{26}$ - $10^{28}$ ergs, generally formed as a result of magnetic reconnection in the chromosphere or above (Cauzzi et al. 2001; Chifor et al. 2008; Archontis & Hansteen 2014; Hong et al. 2016). Various magnetic configurations can lead to this reconnection, with the most prominent cause in the literature being due to emerging magnetic flux interacting with a pre-existing magnetic field structure (Kano et al. 2010; Shimizu 2011; Jiang et al. 2012; Leiko & Kondrashova 2015).

Most micro-flares recorded in the literature appear to be rooted in the chromosphere while showing strong responses in higher energy coronal lines (Chifor et al. 2008; Brosius & Holman 2010; Chen & Ding 2010; Hannah et al. 2011; Gontikakis et al. 2013), most similar in appearance to flares, only on a smaller scale. Due to their appearance in coronal lines, this has indicated upper temperatures for these events of the order of 1 x $10^7$ K (Chifor et al. 2008).

Bi-directional flows at micro-flaring locations are also commonly reported, with typical values ranging from ± 40 - 80 km s$^{-1}$ (Berkebile-Stoiser et al. 2009; Archontis & Hansteen 2014; Chen & Ding 2010; Leiko & Kondrashova 2015; Hong et al. 2016). Numerical simulations of chromospheric micro-flares due to magnetic reconnection appear to also show bi-directional flows either side of the X-point, with the up-flow appearing with a higher velocity than the down-flow (Jiang et al. 2010). Surges from these locations have also been reported (Zuccarello et al. 2011; Jiang et al. 2012).

Ellerman Bombs (EBs; Ellerman (1917)) are localized photospheric reconnection (Georgoulis et al. 2002; Watanabe et al. 2008; Vissers et al. 2013), with flux cancellation rates circa $10^{14}$ – $10^{15}$ Mx s$^{-1}$ (Reid et al. 2016). Their main observational signature is brightening in the wings of the Hα and Ca II 8542 Å line profiles without disturbing the line cores (Rutten et al. 2013; Vissers et al. 2013; Hong et al. 2014). As such, EBs are thought to have no effect on the chromosphere or upper solar atmosphere (Rutten et al. 2013; Vissers et al. 2013), and are considered as solely upper photospheric/lower chromospheric phenomena. Chromospheric surges in Hα have been reported (Yang et al. 2013; Reid et al. 2015), though these are rarely observed, and do not enhance the Hα line core.

Semi-empirical modeling attempts of EBs show local temperature enhancements of ~ 600 - 3000 K at the temperature minimum region (Fang et al. 2006; Berlicki et al. 2010; Berlicki & Heinzel 2014; Li et al. 2015; Grubecka et al. 2016), while recent radiative hydrodynamical modeling appears to show similar results (Reid et al. 2017). Bi-directional flows have also been reported in EBs, with the up-flows showing a higher velocity than the down-flows at EB locations, with values much lower than that of micro-flares (± 5 km s$^{-1}$) (Isobe et al. 2007; Matsumoto et al. 2008; Archontis & Hood 2009; Reid et al. 2017).

A recent study by Hansteen et al. (2017) simulates reconnection events in the solar atmosphere and shows the difference in atmospheric conditions for EBs, micro-flares, and UV bursts. They found EBs to be localised reconnection in the photosphere, with temperatures below 10,000 K, and bi-directional flows of up to ~20 km s$^{-1}$. The micro-flares appeared as reconnection along a polarity inversion line, with temperature enhancements stretching from it's base in the chromosphere to the corona, with values of ~ 1 MK, and bi-directional flows of up to ± 75 km s$^{-1}$. The simulated UV bursts were then a middle ground between these two reconnection topologies, with chromospheric/transition region heating up to $10^5$ K, and flows up to ± 50 km s$^{-1}$.

The similarities between EBs and micro-flares has been noted previously (Berlicki & Heinzel 2014), with Jess et al. (2010) showing evidence of forced reconnection in the photosphere and labelling it as both a micro-flare, and EB. This event also featured bi-directional flows more in-line with EB values, along with Hα wing enhancements.

In this paper, we use the NICOLE inversion code (Socas-Navarro et al. 2015) in an attempt to recover the atmospheric parameters surrounding a chromospheric brightening. Section 2 discusses the observational setup and data reduction. Section 3 discusses the optimisation of the inversions, while Sections 4 & 5 show the results and analysis of the final in-verted data.

## 2. OBSERVATIONS

The observations were carried out with the CRisp Imaging SpectroPolarimeter (CRISP) at the Swedish 1-m Solar Telescope (SST, Scharmer *et al.* 2003; Scharmer *et al.* 2008) on La Palma. The target was active region NOAA 12121, near disk centre (coordinates: X= $40''$, Y= $40''$, $\mu$ = 0.99). The observations took place on 2014 July 28 between 10:43 - 11:24 UT. The observations comprised of imaging spectropolarimetry along the Ca II 8542 Å line. The scan consisted of 15 line positions, taken at ± 0.942 Å, ± 0.580 Å, ± 0.398 Å, ± 0.290 Å, ± 0.217 Å, ± 0.145 Å, ± 0.073 Å, and line center. The data had a post-reduction mean cadence of 30 seconds, and a pixel resolution of $0.059''$/pix. Sample images from the reduced dataset can be seen in Figure 1. The field-of-view (FOV) of the dataset is 59 x $58''$.

The data was processed with the Multi-Object Multi-Frame Blind Deconvolution (MOMFBD) algorithm (Löfdahl 2002; van Noort *et al.* 2005). This includes tessellation of the images into 64x64 pixels$^2$ sub-images for individual restoration to preserve the assumption of invariance of the image formation models. An extended MOMFBD scheme that includes the reconstruction of auxiliary wide-band images, co-temporal with the narrow-band wavelengths and polarizer states, were used together with de-stretching (Shine *et al.* 1994) to reduce the impact of residual seeing on the profiles and on cross talk (Henriques 2012). Prefilter field-of-view and wavelength dependent corrections were applied to the restored images. Further information on the general MOMFBD pipeline is available in de la Cruz Rodríguez *et al.* (2015).

A brightening can be seen in the line core just below the sunspot in Figure 1, as well as in the line wings, where it appears most prominently. This brightening has small footpoints in the far wings of the Ca II 8542 Å observations, where the photosphere is sampled. This brightening therefore stretches from the photosphere up to the mid chromosphere, with its main enhancements occurring below the canopy. Some small sub-structure can also be seen in the brightening, showing that the initial ball of emission appears to split into two brightened regions. Both of these regions appear to visibly connect to a dark overlying chromospheric fibril (best seen in the -0.290 Å images of Figure 1 or Movie1).

Co-aligned Solar Dynamics Observatory (SDO) Atmospheric Imaging Assembly (AIA, Lemen *et al.* 2012) and Helioseismic and Magnetic Imager (HMI, Schou *et al.* 2012) data were also created, in an attempt to locate the same brightening in the transition region or coronal lines. A minor enhancement is apparent in the 1600 Å and 1700 Å channels of SDO, but no upper atmospheric response to this event was visible. An overlying filament is apparent in the upper atmospheric lines of AIA (especially He II 304 Å), and may block any emission of the brightened/ event below.

## 3. NICOLE INVERSIONS

NICOLE is a parallelized code that solves multi-level, non-LTE (NLTE) problems following the preconditioning approach described in Socas-Navarro & Bueno (1997), while also taking into account polarisation of the light via Zeeman splitting. NICOLE uses response functions combined with standard fitting techniques to obtain an atmospheric model which best describes the observed Stokes profiles (Socas-

Navarro *et al.* 1998). The inversions require an initial model to be perturbed, which contains parameters such as a temperature profile, line-of-sight velocity, magnetic field vector, electron density, and micro-turbulence.

The electron and gas pressures are attained from inserting the temperature stratification into an equation of state with hydrostatic equilibrium imposed and an upper boundary in electron pressure. Isotropic scattering and complete frequency redistribution (CRD) are also assumed by NICOLE. While the resonant lines of Ca II H & K require partial redistribution to be fully modeled, the Ca II 8542 Å line can be accurately synthesized in CRD. NICOLE operates in 1.5D, whereby each pixel is individually inverted with its own independent atmosphere. While NICOLE uses a plane parallel atmosphere, radiation comes from and scatters to all directions (I- and I+), with each direction seeing a different effective atmosphere. The correct radiation field is important when computing the NLTE populations of the different levels. NICOLE supports up to five angles along a Gaussian quadrature (see e.g. section 5.1.2 of Rutten (2003) for further details of such numerical approximation in this context). We selected 3 angles which is a common compromise between accuracy and speed.

The Ca II 8542 Å line transition occurs between a 4P - 3D shell, with an effective landé factor of $g_{eff}$ = 1.1. The nuclear angular momentum of the Ca II atom is non-zero ($F \neq 0$), and as such, small hyperfine structure will also be apparent in the resultant Zeeman-induced profiles, and must be considered within the inversion code (de la Cruz Rodríguez *et al.* 2012). The NICOLE atomic and line transition data for Ca II 8542 Å were added into their associated lookup libraries within NICOLE, along with the 5 isotopes of varying nuclear angular momentum.

Due to the contribution function of Ca II 8542 Å encompassing a broad range of heights along the photosphere and chromosphere (Cauzzi *et al.* 2008; Reid *et al.* 2017), some stratification with height would also be necessary in order to fully utilize the observations.

The input observations were optimized in order to accumulate sufficient signal in Stokes-$Q$, $U$, and $V$. The area within the red box of Figure 1 was used for the inversions, and was spatially binned 2x2. The profiles were normalized to a reference profile, taken from areas of quiet-Sun over the whole FOV, averaged over the whole time series. The profiles were then interpolated onto a wavelength grid with 36 mÅ spacing between points. The weights for Stokes-$Q$ and $U$ were made to be half that of Stokes-$I$ and $V$. Non-observed points in the interpolated profiles are given a negligible weight.

In order to improve convergence while maintaining many degrees of freedom, a staggered approach is taken, similar to that discussed in Ruiz Cobo & del Toro Iniesta (1992). The staggered approach uses an initial inversion considering a lower number of nodes to get to a reasonable answer, albeit with little vertical stratification. The output model from this can then be used as the input guess model in a second inversion with more nodes.

The initial inversions, named '4 node' inversions, consisted of 4 nodes in temperature, 3 nodes in line-of-sight velocity, 1 node in line-of-sight magnetic field, and 1 node in microturbulence. An example output frame of this can be seen in Figure 2. The fitted Stokes-$I$ line profile is quite good, with the temperature profile mapping the brightened chromospheric structure accurately. The Stokes line profiles shown in panels 3 and 4 of Figure 2 are taken from a pixel at the base of the con-nected fibril brightening with the microflaring structure. By only allowing 1 value of magnetic flux density over the formation height of the whole line, the synthetic profiles cannot map depth dependent magnetic flux regions. There appears to be a temperature enhancement throughout the brightened area. The line-of-sight velocity appears to show bi-directional flows at the micro-flare location, up to 20 km s$^{-1}$, mainly in the chromosphere. The microturbulence appears to be quite large (up to ~ 20 km s$^{-1}$) in some unsuccessfully inverted regions, and around 10 km s$^{-1}$ in the brightened region. This implies that with more nodes, better fitting could be achieved in the temperature, velocity, and magnetic flux density, reducing the amount of unresolved microturbulence.

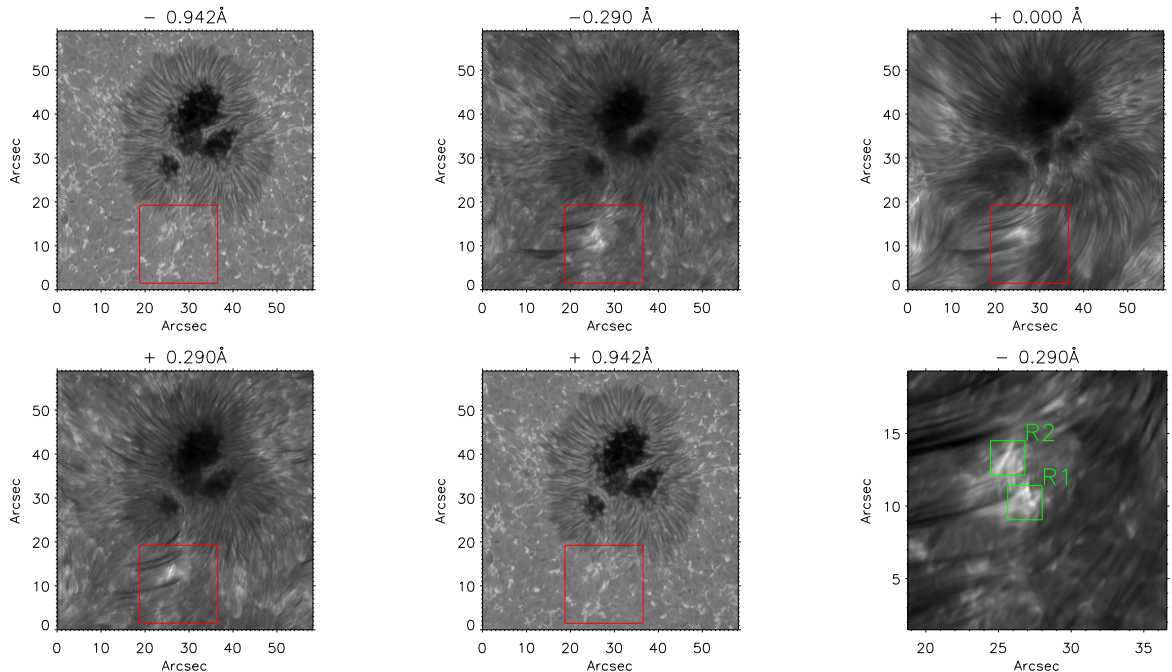

**Figure 1.** Images at various wavelengths across the Ca II 8542 Å line profile, showing the micro-flare event and connected fibril. The red box highlights the area shown in the bottom right panel. The green boxes show the bright regions (labeled as R1 and R2). A movie of the bottom right panel is available (Movie 1).

This output model was then smoothed before being fed into a new inversion with more nodes. This process spatially smooths the changes in the atmosphere introduced by the code by measuring the difference between the output model and the initial guess model from the 4 node inversion. This smoothed difference is then reapplied to the initial guess model (following de la Cruz Rodríguez, private communication). The new inversion uses 7 nodes in temperature (named '7 node'), in an attempt to identify local fine thermal stratification in the atmosphere. Five nodes in line-of-sight velocity are also used, with 3 nodes in line-of-sight magnetic field, to allow for different flux densities in the photosphere than the chromosphere. One node is also used in the transverse magnetic field, though the weights are lessened here due to the noisiness of the Stokes-$Q$ and $U$ profiles. Again, 1 node is applied in microturbulence. In this inversion, the regularisation is reduced slightly to allow for increased vertical stratification.

The output for the 7 node inversion can be seen in Figure 3. The fitting of the line profiles is better than from the 4 node. There appears to be very little transverse magnetic fields present in the micro-flaring region, similar to that reported for EBs in Reid *et al.* (2016). The line-of-sight magnetic field appears to weaken in the higher atmospheric heights, as would be expected with magnetic flux densities between the photosphere and the chromosphere. The thermal enhancement still appears to exist, with higher contrast in the brightened region in the chromosphere. The line-of-sight velocity fields still show the bi-directional Doppler shifts, and look very similar to the 4 node output. However, the down-flow related to the connected brightened fibril structure is now apparent. The micro-turbulence appears to be much weaker in the 7 node inversion, reaching no more than 3.5 km s$^{-1}$, completely removing the very high, saturated regions (20 km s$^{-1}$) present in the 4 node inversions. This is due to the increased temperature nodes allowing for a better fitting of the thermal broadening.

## 4. INVERSION RESULTS

The inversions were ran in steps of 2 scans (~60 seconds). Initially it was thought that by setting the initial guess model for an inversion to be the smoothed output model from the previous frame, a better convergence could be achieved. However, it appeared to only restrict the possible outcomes in the dynamic event. With 60 seconds between frames, the scene can vary largely, and the best contrast and fitting of the line profile was found when the 4 node inversions for all frames used the same initially inverted output which used the FALC model as the initial guess. This initial inversion was done for the median time frame during the lifetime of the event. The microturbulence was reset and the output of this inversion was also smoothed spatially to avoid excess inversion noise.

The outputs of the inversions can be seen in Figure 3 for T = 1020s, while Figure 4 shows the inversion outputs over time. Movie 2 also presents detailed inversion outputs. R1 appears to be heated first, with a morphology similar to that of Ellerman Bombs (Watanabe *et al.* 2011), while R2 appears to get heated later in the timeseries. Both regions appear to be connected to a dark chromospheric fibril structure. Figure 4 shows that as R2 evolves and becomes stronger, the length of heating along the fibril also increases. The propagation of the heating along the fibril appears to occur at a rate of ~6 km s$^{-1}$.

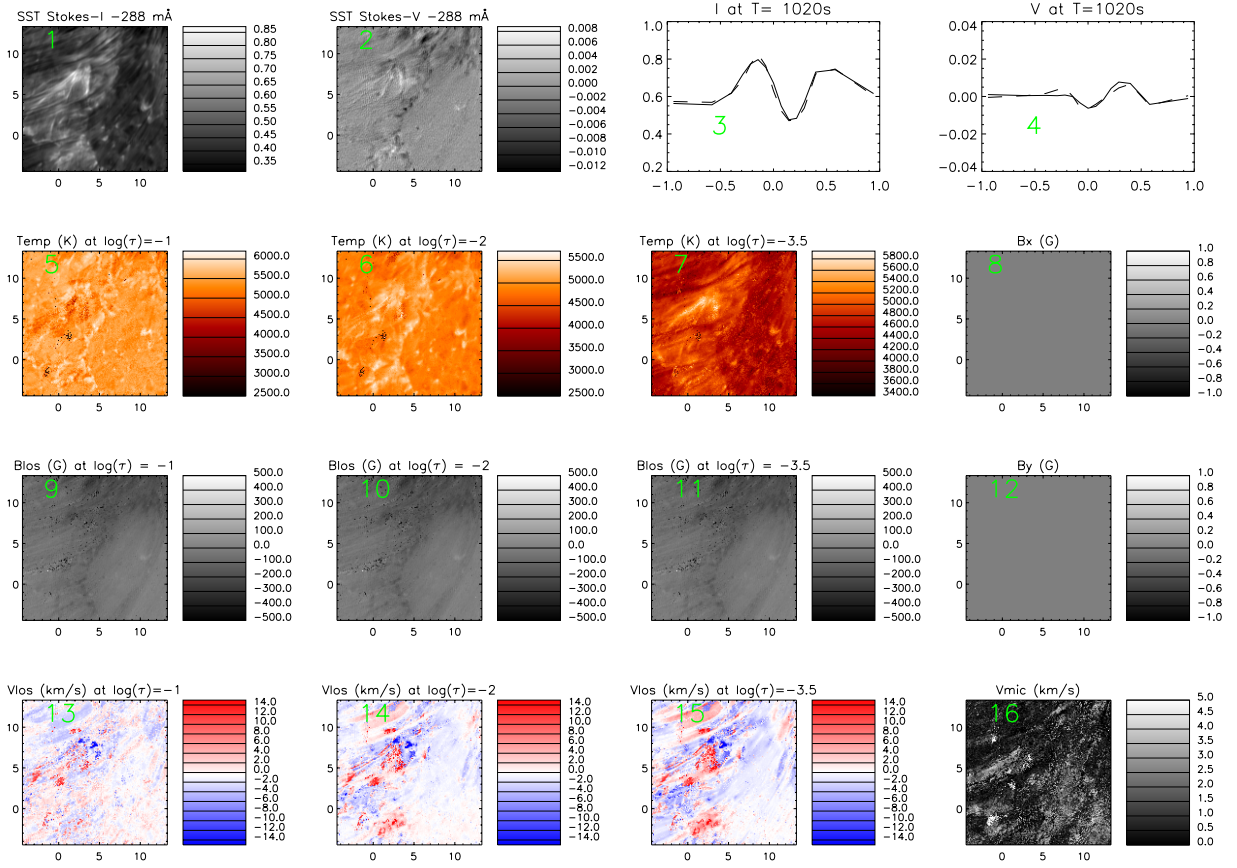

**Figure 2.** Panel 1: The inverted Stokes-*I* observations at -288 mÅ from line core. Panel 2: Co-spatial Stokes-*V* observations in -288 mÅ from line core. Panel 3: The Stokes-*I* line profiles of the observations (solid) and the synthetic, fitted profiles from the inversions (dashed). The line profiles are taken from a pixel within the brightening. Panel 4: The Stokes-*V* line profiles, co-spatial to Stokes-*I*. Panels 5-7: The NICOLE output model showing the temperatures at various optical depths. Panel 8: A transverse component of magnetic field. Panels 9-11: The line-of-sight magnetic flux density at various optical depths. Panel 12: The secondary transverse component of magnetic field. Panels 13-15: Line-of-sight velocity at various optical depths (positive = down-flow). Panel 16: Microturbulent velocity (km s$^{-1}$).

Firstly, the temperature profile was investigated. A 20 binned pixel$^2$ box was placed around the brightened footpoints (R1 and R2). The average temperature profile could be taken for successfully inverted pixels to attain how the average temperature in the micro-flaring regions change over time. Figure 5 shows the average temperature profile compared to the background profile. The BG AVG profile (black line) is taken as the background average of those pixels not affected by the flaring event, while the MF AVG profile (pink line) is the average taken from the micro-flaring region. Three individual pixels from across the brightenings are also shown to present the temperature profiles for successfully inverted pixels. One of the pixels from the 4 node inversion is also shown (LN).

The average profile shows temperature rises in the chromosphere and photosphere, with a slight gradient. When looking at individual pixels, it appears as though NICOLE found its best fits when inserting chromospheric temperature bumps around $log(\tau) = -3.5$. The temperature outputs were then averaged between $log(\tau) = -3$ and $log(\tau) = -4$, and the temperature enhancements relative to the background average calculated. Figure 6 shows the relative temperature enhancements for both micro-flaring regions. Region 2 shows more consistent, stronger temperature enhancements throughout the event, while at T = 850s, Region 1 reaches the end of it's lifetime. The mean temperature enhancements for Regions 1 and 2 are 400 K and 750 K during their lifetimes respectively, though the central pixels chosen in Figure 5 show that the temperature enhancements are concentrated in the brightening and can reach 2000 K.

Figures 2 and 3 appear to show strong bi-directional Doppler shifting in the local area during the event. To check the Doppler shifting in the event over time, 1 km s$^{-1}$ bins were created, up to ± 25 km s$^{-1}$. A box was then placed around R1 and R2, and each pixel placed into a bin. This was done over all inverted frames, at 3 different optical depths, again averaged over 1 dex. Figure 7 shows the output from this in both regions from the 7 node outputs. Region 1 shows strong chromospheric red-shifting throughout the event, which is concentrated towards the end of the observations. Region 2 however appears to show a near equal amount of red and blue-shift in the chromosphere. Figure 3 shows that while strong blue shifting appears around the heated region (R2), the heated area, which follows the connected fibril, appears redshifted. The low sensitivity to the photosphere which is sampled in the line wings is apparent at $log(\tau) = -1$. The extrapolation of the velocity estimates from the chromosphere pushes some of the pixels to extreme ± 20 km s$^{-1}$ values.

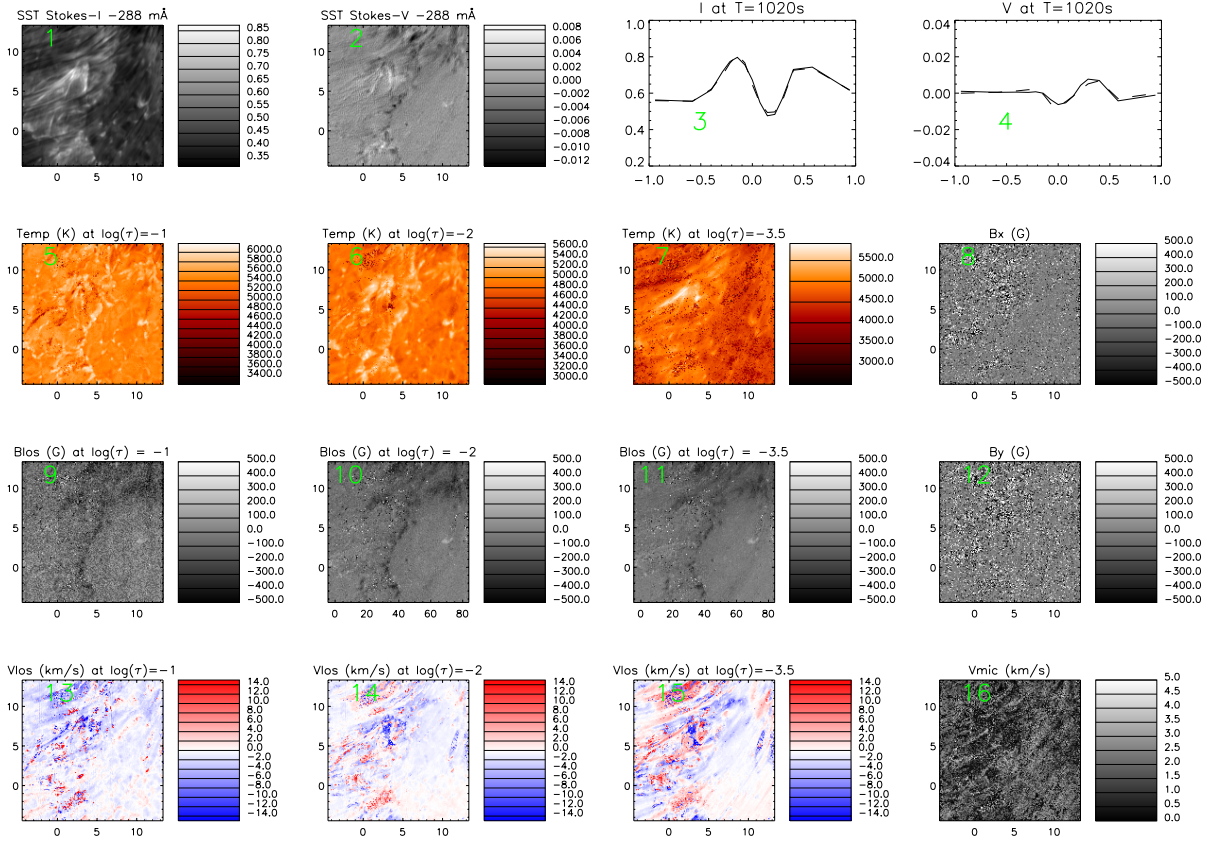

**Figure 3.** Inversion outputs for the 7 node inversion. The description of the panels is identical to those in Figure 2. Movie 2 shows the temporal evolution from the inversion outputs.

Centre of gravity (COG) measurements (Uitenbroek 2003) were also used to determine the Doppler shifting of the observed line profiles at the heated fibril location. The results match the inversion outputs, showing a red-shift of the heated region of the fibril (∼ up to 15 km s$^{-1}$). Figure 8 shows the observations of the fibril region along with the relative line-of-sight velocity measurements from the COG method.

This strong down-flow measurement appears to continue throughout the brightened region of the fibril, while the rest of the fibril is only slightly red-shifted. The brightened regions of R2 are constantly red-shifted rather strongly, while this effect is more subtle with R1.

The magnetic flux was also investigated for the 2 regions. Using the same box as for the velocity and temperature measurements, the flux evolution is plotted for the 2 regions and for the different polarities in Figure 9. These values are from the 7 node inversions and averaged over 1 dex, centered around $log(\tau) = -3.5$. The flux cancellation values can be seen in Table 1.

The overall flux for the 4 node output in R1 is lessened, with the majority of this flux cancellation occurring in the negative polarity, with the positive polarity fields appearing negligible. The 4 node output for R2 tells a different story. Rather than showing evidence of flux cancellation, the opposite is apparent. Again, this is predominately due to the negative magnetic flux, with the positive polarity flux difference being 5 orders of magnitude less.

In the 7 node inversions, the magnetic output from the inversions contained some depth with 3 nodes in line-of-sight magnetic flux density. At $log(\tau) = -3.5$, R1 appears still to have strong magnetic flux cancellation. This is again mainly due to the negative polarity flux difference, with the positive polarity flux cancellation appearing stronger than that of the 4 node outputs. The overall flux cancellation rate is strongest around $log(\tau) = -3.5$, with weaker values in the photosphere below. R2 again shown flux growth due to strong enhancement in the negative polarity flux. The positive polarity flux did show some weak flux cancellation in the 7 node outputs at this optical depth. Again, these values are weakened at higher optical depths.

## 5. DISCUSSION AND CONCLUSIONS

We have shown evidence of a localized chromospheric brightening with foot-points extending down to the upper photosphere, connecting to a chromospheric fibril. Using NICOLE inversions, we were able to show localized temperature enhancements in the low-mid chromosphere of the order of ~1000 K, with no upper atmospheric response apparent. However, an overlying filament structure, which is present in the SDO AIA data, could mask any signal of this event at higher atmospheric heights. The temperature values are very mild in comparison to the previous literature on micro-flares. The localized temperature enhancement bumps are very similar to that shown in previous modeling efforts of Ellerman Bombs (e.g. Berlicki & Heinzel (2014); Reid et al. (2017)), in shape and magnitude, only occurring slightly higher in the atmosphere, thus disturbing the Ca II 8542 Å line core formation in the mid-chromosphere. The heated region also appeared to have the electron density increase tenfold (up to $10^{13}$ cm$^{-3}$ in the low-mid chromosphere). The localized temperature enhancements and velocity values are most similar to that of EBs in Figure 7 of Hansteen et al. (2017), not reaching the strong temperatures and velocities of their UV bursts and small-flare examples which stretch into the transition region and corona.

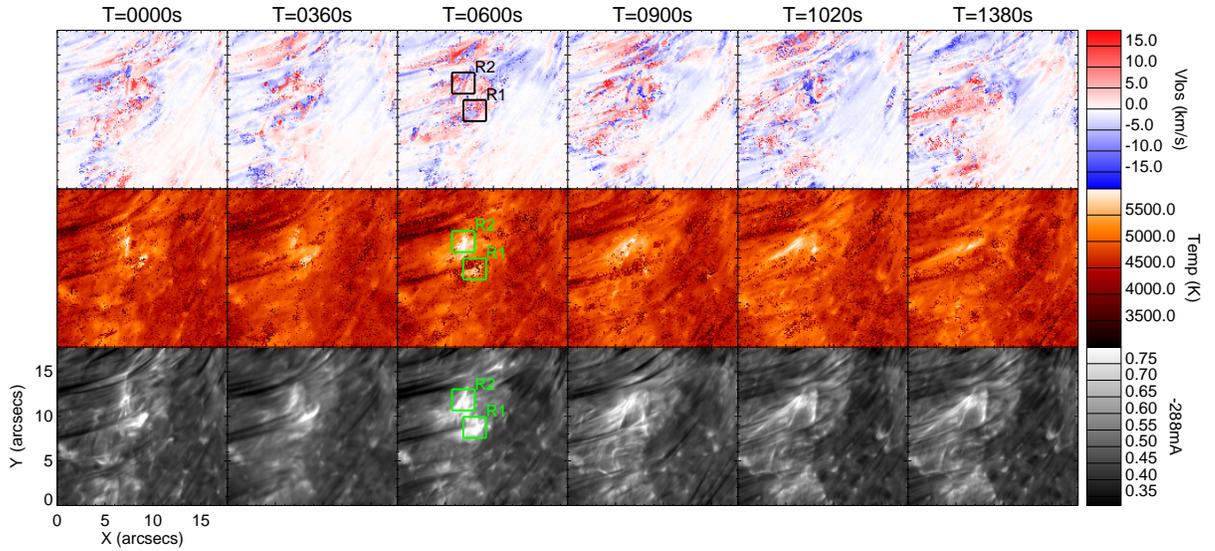

**Figure 4.** Top: The line-of-sight velocity outputs averaged between $log(\tau) = -3$ and $log(\tau) = -4$. Middle: The temperature output averaged over the same optical depth range. Bottom: The observed Stokes-$I$ -288 mÅ images. The FOV is identical to the red box of Figure 1.

|  | R1 (4 node) | R2 (4 node) | R1 (7 node) | R2 (7 node) |
|---|---|---|---|---|
| **Overall** | $(8.89 \pm 1.90) \times 10^{16}$ | $(-4.02 \pm 0.92) \times 10^{16}$ | $(1.57 \pm 0.32) \times 10^{17}$ | $(-9.73 \pm 2.04) \times 10^{16}$ |
| **Positive** | $(1.97 \pm 0.42) \times 10^{11}$ | $(-5.76 \pm 1.38) \times 10^{11}$ Mx s$^{-1}$ | $(5.19 \pm 1.30) \times 10^{15}$ | $(2.39 \pm 0.55) \times 10^{15}$ |
| **Negative** | $(8.80 \pm 1.75) \times 10^{16}$ | $(-3.98 \pm 0.76) \times 10^{16}$ | $(1.42 \pm 0.31) \times 10^{17}$ | $(-9.97 \pm 2.39) \times 10^{16}$ |

**Table 1**
Magnetic flux cancellation measurements made for the brightened regions (R1 and R2), for the 4 node and 7 node outputs. The overall flux cancellation is shown, alongside the positive and negative polarity flux cancellation values. Values shown are in Mx s$^{-1}$.

Two regions were identified as the foot-points of this event, both showing temperature increases and connections to the overlying fibril. Weaker temperature enhancements were apparent in one of the foot-points (R1) than it's counterpart (R2). Figure 4 appears to show strong overlying chromospheric canopy at the location of R1 (most apparent in T=600s). This reduction in intensity in the core and near wings of the Ca II 8542 Å line reduces the strength of the temperature response function for the chromosphere, resulting in weaker temperature enhancements. Instead, the cooler overlying structure is identified and masks the event. The temperature enhancements presented from the inversions should then be treated as a lower boundary.

Flux cancellation was apparent from the outputs of the inversions for one of the two heated foot-points studied (R1). This is stronger than that of typical EBs or moving magnetic features (MMFs) (Reid et al. 2016; Nelson et al. 2016). R2 however appeared not to show evidence for flux cancellation, perhaps due to the more complex nature of the event and its stronger connection to the overlying fibril. No strong change in magnetic flux was observed via the line-of-sight magnetograms from HMI. This may be due to the low formation height of the 6173 Å line sampled by HMI, or the relatively low resolution of the instrument in comparison to ground based instrumentation.

Initially, R2 had a relatively weak Ca II 8542 Å line enhancement in comparison to it's eruptive counterpart (R1). However, as the brightening of R2 changed morphology and it connected to the chromospheric fibril, very strong, impulsive heating occurred at the base of this connection (up to ~2000 K). The heating progresses along the fibril at a rate of ~6 km s$^{-1}$. This is similar to the extension speed of the Ellerman Bomb 'flame' morphology in the photosphere (Watanabe et al. 2011).

The heated regions connecting to the fibril also show apparent downward, red-shifted motion of the order of 10 - 15 km s$^{-1}$. Figure 3 hints that there may be up-flowing material be-low $log(\tau) = -3.5$. This may interact with the chromospheric canopy to form this brightening. If this was an upwardly moving positive polarity magnetic flux region, this could be evidence of a form of chromospheric reconnection similar to previous micro-flares and to mechanisms describing Ellerman Bomb formation (Georgoulis *et al.* 2002). This would also explain the chromospheric flux reducing for the negative polarity, but not for the positive polarity, as the positive polarity flux region moved into this portion of the atmosphere. However, due to the weaker contribution of the Ca II 8542 Å line to the photosphere compared to the chromosphere, combined with the low signal from circular polarisation, obtaining accurate photospheric magnetic fields would require co-temporal spectro-polarimetry from another spectral line. In this particular case there seems to be a pre-existing, overhanging canopy (e.g. see beginning of Movie1 for the best impression), of which part then connects to the lower EB flame. The down-flow in the upper fibril (connecting to R2) could indicate a draining of the pre-existing canopy post reconnection.

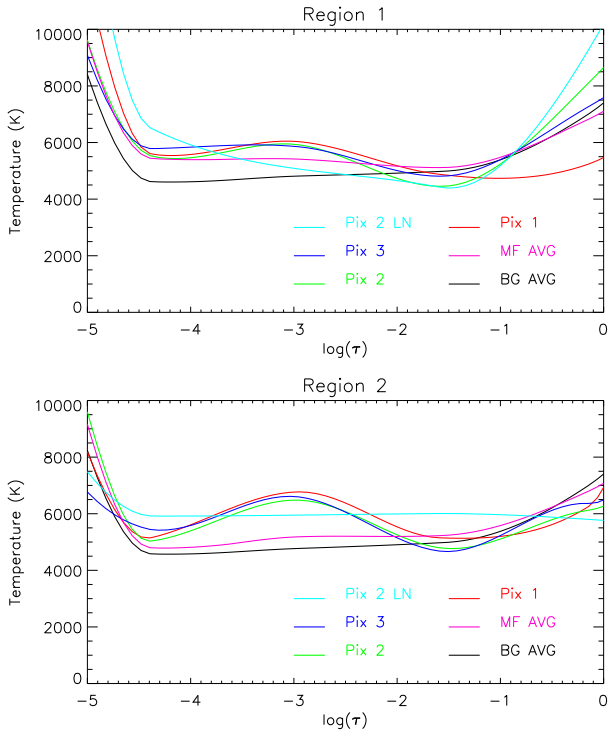

**Figure 5.** Temperature profiles of various pixels within R1 and R2, compared to the region average (MF) and background average (BG). Pix 2 LN shows the 4 node output for Pix 2.

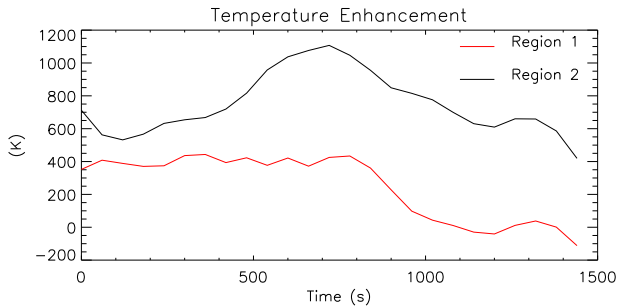

**Figure 6.** Mean temperature enhancements of the micro-flaring regions over time. This was measured over a slab between $log(\tau) = -3$ and $log(\tau) = -4$.

While strict nomenclature exists to differentiate between a plethora of solar events, we postulate that the difference between this event (named as a micro-flare due to its chromospheric signatures) is identical to that of an Ellerman Bomb, with the only difference being that this event is slightly higher in the atmosphere than a traditional EB, causing the moustache line profile normally associated with EBs to be skewed. Due to this event occurring slightly higher in the atmosphere than typical Ellerman Bombs, the usual EB 'flame' morphology interacts with the chromospheric canopy, suggesting that not all EB reconnection events are purely photospheric in nature.

Armagh Observatory and Planetarium is grant-aided by the N. Ireland Department for Communities. The Swedish 1-m Solar Telescope is operated on the island of La Palma by the Institute for Solar Physics of Stockholm University in the Spanish Observatorio del Roque de los Muchachos of the Instituto de Astrofísica de Canarias. The authors wish to acknowledge the DJEI/DES/SFI/HEA Irish Centre for High-End Computing (ICHEC) for the provision of computing facilities and support. AR would like to thank Armagh Observatory and Queen's University Belfast for funding. The research leading to these results has received funding from the European Community's Seventh Framework Programme (FP7/2007-2013) under grant agreement no. 606862 (F-CHROMA). Research at the Armagh Observatory is grant-aided by the Northern Ireland Department of Communities.

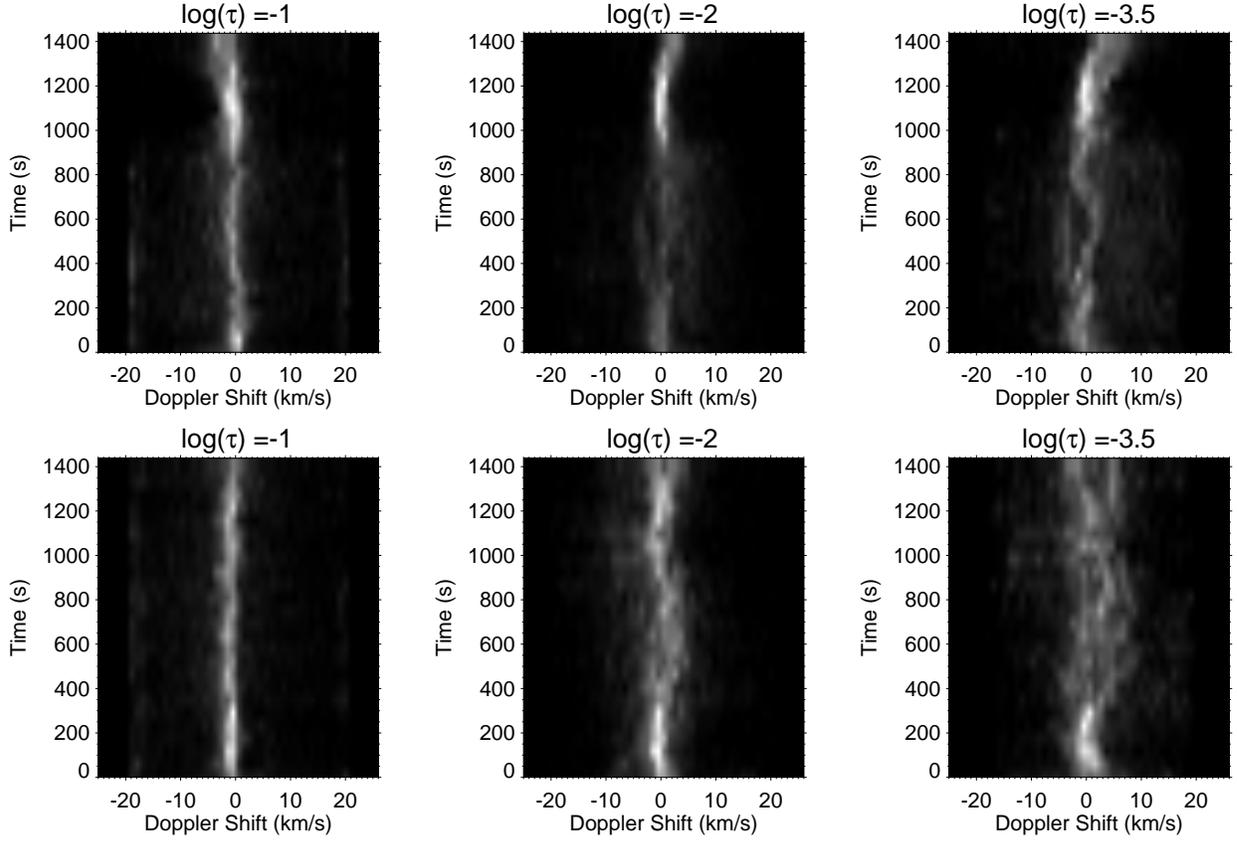

**Figure 7.** Velocity maps showing the Doppler shifts over time. Brightness indicates pixel density within the bin. Top row shows R1, while the bottom row shows R2.

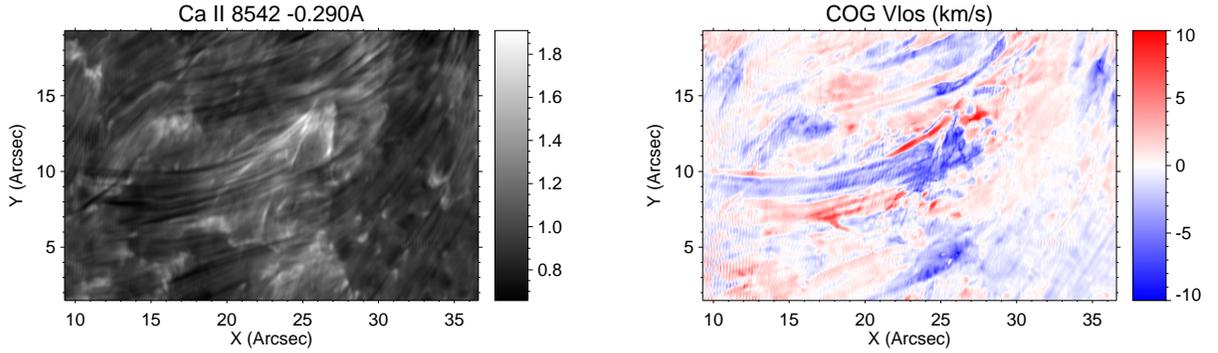

**Figure 8.** Left: The observed micro-flaring region with the associated fibril. Right: The line-of-sight velocity obtained via the centre of gravity method. The frame chosen is identical to that of Figure 2 and Figure 3.

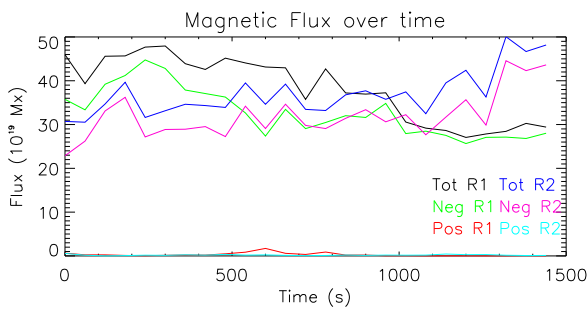

**Figure 9.** Magnetic flux over time. 'Tot' = Total Flux, 'Pos' = positive flux only, and 'Neg' = negative flux only. R1 and R2 refer to the areas shown in the bottom right panel of Figure 1.